\documentclass[aps,eqsecnum,nofootinbib,superscriptaddress,showpacs]{revtex4}
\usepackage{amssymb, amsmath, amsopn, amsthm}
\usepackage{epsfig,amsfonts,bm}
\usepackage{graphicx}
\newcommand{\normrho}[1]{\overline{#1}} 

\begin{document}
\title{A holographic superconductor model in a spatially anisotropic background}
\date{\today} 
\begin{abstract}
We investigate an anisotropic model of superconductors in the Einstein-Maxwell-dilaton theory 
with a charged scalar field. It is found that the critical temperature decreases as the anisotropy becomes large. We then estimate the energy gap of the superconductor, 
and find that the ratio of the energy gap to the critical temperature 
increases as the anisotropy increases and so it is larger than that in the isotropic case. 
We also find that peudogap appears due to the anisotropy. 
\end{abstract}
\author{Jun-ichirou Koga}
\email{koga@waseda.jp}
\affiliation{Research Institute for Science and Engineering, Waseda University, Shinjuku, Tokyo 169-8555, Japan}
\affiliation{Faculty of Engineering, Shibaura Institute of Technology, Saitama, 330-8570, Japan}
\author{Kengo Maeda} 
\email{maeda302@sic.shibaura-it.ac.jp}
\affiliation{Faculty of Engineering, Shibaura Institute of Technology, Saitama, 330-8570, Japan}
\author{Kentaro Tomoda} 
\email{k-tomoda@kwansei.ac.jp}
\affiliation{Department of Physics, Kwansei Gakuin University, Sanda, Hyogo, 669-1337, Japan}


\maketitle

\section{Introduction}\label{sec:intro}
The AdS/CFT correspondence~\cite{adscft} gives new insight 
for understanding strongly coupled gauge theories or strongly 
correlated condensed matter systems. 
In particular, high temperature superconductors in the framework of the AdS/CFT correspondence attract much attention. 
The simple model of holographic superconductors initiated in \cite{HHHletter2008, HHH2008} 
has been extended into more realistic models with inhomogeneity \cite{MaedaOkamuraKoga, 
HorowitzSantosTong2012_1, HorowitzSantosTong2012_2, HorowitzSantos2013}. 
Anisotropic models of superconductors also have been investigated 
in the context of p-wave superconductors, 
where a non-Abelian gauge field condenses in the superconducting state \cite{GubserPufu2008, RobertsHartnoll2008, HutasoitSiopsisTherrien2014}. 
Since real-world superconductors exhibit various types of anisotropy, 
it will be valuable to consider other models of anisotropic holographic superconductors. 

We then consider in this paper the holographic model 
in the Einstein-Maxwell-dilaton theory analyzed 
in \cite{IizukaMaeda2012}, where the bulk solution corresponding to the normal state 
was constructed. 
This work was motivated by the IR geometry, 
i.e., the near-horizon geometry, of an asymptotically anti-de Sitter~(\emph{AdS}) black \emph{brane} spacetime, 
since the behavior of a holographic superconductor strongly depends on the IR geometry, 
as we see in the analyses of the Fermi surface of a non-Fermi liquid \cite{IKNT2012}. 
While anisotropy in the IR geometry has been actively investigated also in the context of 
Lifshitz geomerty \cite{KachruLiuMulligan2008,Taylor2008,GKPT2010,GIKPTW2010}, 
where anisotropy between time and space is considered,  
the anisotropy in \cite{IizukaMaeda2012} is between 
two spatial directions induced by spatial dependence 
of the dilaton field, i.e., the Bianchi type \cite{IKKNST2012}. 

In this paper, we construct the superconducting state in the Einstein-Maxwell-dilaton theory 
by turning on a charged scalar field, 
and analyze the effect of the anisotropy on the optical conductivity. 
It will be worthwhile mentioning here the difference between our model and the p-wave holographic model of superconductors \cite{GubserPufu2008, RobertsHartnoll2008, HutasoitSiopsisTherrien2014}.   
In our model, the bulk solution is anisotropic not only in the superconducting state but also in the normal state, 
as if the anisotropy arose due to structures of a superconductor, such as 
crystal structure and doping. In contrast, the anisotropy from the non-Abelian gauge field vanishes 
in the normal state in the p-wave model of superconductors \cite{GubserPufu2008, RobertsHartnoll2008, HutasoitSiopsisTherrien2014}. 
We also point out that, in our model, the anisotropy is described by a parameter 
in the \emph{solution} of the dilaton field and hence it is under our control, 
while the parameter of the anisotropy in \cite{GubserPufu2008, RobertsHartnoll2008, HutasoitSiopsisTherrien2014} is a coupling constant of the \emph{theory}. 
Therefore, it seems natural to view our model as corresponding to 
a superconductor with the anisotropy due to crystal structure and doping. 

We numerically construct four-dimensional black brane background bulk solutions dressed with a charged scalar field 
and investigate the properties of the superconducting states in Sec. \ref{sec:background}. 
We then perturb the solutions by the gauge field and investigate the optical conductivities in Sec. \ref{sec:conductivity}. 
From the low frequency behavior we will determine the energy gap. 
Conclusion and discussions are 
devoted in Sec. \ref{sec:discussion}. 

\section{Backgrounds}\label{sec:background}
In this section, we numerically construct anisotropic black brane solutions 
in the Einstein-Maxwell-dilaton theory coupled to a charged scalar field $\Psi$. 
The solutions with a charged scalar hair correspond to a superconducting state 
in the dual field theory. When $\Psi$ is identically zero, 
they are reduced to the solutions constructed in \cite{IizukaMaeda2012},  
which correspond to a normal state in the dual field theory. 
Not only the normal state solutions, we construct here the superconducting state solutions.

\subsection{Preliminaries}\label{subsec:pre}
The action we consider is
\begin{eqnarray}
S &=& \int d^4 x \sqrt{- g} \left[ R + \frac{6}{L^2} - \frac{1}{4} F^{a b} F_{a b} - ( \nabla^a \varphi ) ( \nabla_a \varphi )
- ( D^a \Psi )^* ( D_a \Psi ) + \frac{2}{L^2} \Psi^* \Psi  \right] , \notag \\
D_a &\equiv & \nabla _a - i q A_a, \qquad F_{a b} \equiv \partial _a A_b - \partial _b A_a,
\label{eqn:action}
\end{eqnarray}
where $\varphi$ is dilaton, $L$ is the AdS length scale and an asterisk denotes complex conjugate.
The field equations derived from the action 
\eqref{eqn:action} take the following form
\begin{eqnarray}
R_{a b} - \frac{1}{2} g_{a b} R - \frac{3}{L^2} g_{a b} &=& T_{a b}, \label{eqn:einstein} \\
\nabla ^a F_{a b} - i q \left( \Psi ^* D _b \Psi - \Psi D ^* _b \Psi ^* \right) &=& 0, \label{eqn:maxwell} \\
D ^a D _a \Psi + \frac{2}{L^{2}} \Psi &=& 0, \label{eqn:complex} \\
\Box \: \varphi &=& 0, \label{eqn:scalar}
\end{eqnarray}
where
\begin{eqnarray}
  T_{a b} &=& \frac{1}{2} F_{a c} F_b^{\ c} - \frac{g_{a b}}{8} F^2 + \nabla _a \varphi \nabla _b \varphi - \frac{g_{a b}}{2} \left( \nabla \varphi \right) ^2 \notag \\
                 && + \frac{g_{a b}}{L^2} |\Psi |^2 - \frac{g_{a b}}{2} |D \Psi |^2 
                  + \frac{1}{2} \left(D _a \Psi D ^*_b \Psi ^* + D _b \Psi D ^*_a \Psi ^* \right). \label{eq:energy-momentum}
\end{eqnarray}

In order to make the definitions of physical quantities clear, here we start with the ansatz of the metric 
\begin{equation}
d s^2 = - \frac{r^2}{L^2} \: g(r) \: d t^2 + \frac{L^2}{r^2} \: g^{-1}(r) \: d r^2 
+ r^2 \left( e^{A(r) + B(r)} \: d x^2 + e^{A(r) - B(r)} \: d y^2 \right) ,  
\label{eqn:MetricOr} 
\end{equation}
together with the forms of the gauge potential $A_a$ and the charged scalar field $\Psi$ as
\begin{equation}
A_a = \phi (r) \; ( d t )_a  ,\quad \Psi = \Psi (r) ,
\label{eqn:GaugeOr} 
\end{equation}
which we assume describes 
an asymptotically AdS black brane spacetime. 
Then, we should have the asymptotic forms of the metric functions as
\begin{equation}
g(r) \rightarrow 1 , \quad A(r) \rightarrow 0 , \quad B(r) \rightarrow 0  \quad 
 \mathrm{as} \quad r \rightarrow \infty .
\end{equation}

The asymptotic behavior of the gauge potential and the charged scalar field is shown to take the form
\begin{eqnarray}
\phi (r) &\rightarrow & \mu - \frac{\rho}{r}  
\quad \mathrm{as} \quad r \rightarrow \infty , \label{eqn:gauge_asympt_r}
\\
\Psi (r) &\rightarrow & \frac{\Psi_1}{r} + \frac{\Psi_2}{r^2} 
\quad \mathrm{as} \quad r \rightarrow \infty , \label{eqn:scalar_asympt_r}
\end{eqnarray}
where $\mu$, $\rho$, $\Psi_1$ and $\Psi_2$ are constants. 
The constants $\mu$ and $\rho$ are the chemical potential and the charge density 
in the dual field theory, respectively. 
On the other hand, for the charged scalar field with the potential given in Eq. \eqref{eqn:action}, 
either constant $\Psi_1$ or $\Psi_2$ corresponds to the expectation value of an operator $\mathcal{O}$ in the dual field theory 
and the other should vanish \cite{HHHletter2008}. 

With the time coordinate $t$ in Eq. \eqref{eqn:MetricOr}, the timelike Killing vector $\xi^a$ is 
expressed as $\xi^a = \left( \partial / \partial t \right)^a$ and then the surface gravity $\kappa$ 
is defined by $\xi^a \nabla_a \xi^b = \kappa \: \xi^b$, with which the Hawking temperature 
$T$ is given  by $T = \kappa / 2 \pi$.  

Now we transform to the coordinate system employed in \cite{IizukaMaeda2012}, 
i.e., we introduce the new radial coordinate $z$ defined as
\begin{equation}
z = \frac{r_+}{r} , 
\label{eqn:ZDef} 
\end{equation}
where $r_+$ is the horizon radius, and we rescale the coordinate variables $( t , x , y )$ in Eq. \eqref{eqn:MetricOr} as
\begin{equation}
\frac{r_+}{L^2} \: t \rightarrow t , \quad 
\frac{r_+}{L} \: x \rightarrow x, \quad 
\frac{r_+}{L} \: y \rightarrow y .
\label{eq:r-z}
\end{equation}
Thus, the metric and the gauge potential are rewritten as
\begin{eqnarray}
d s^2 &=& \frac{L^2}{z^2} \left[ - \: g(z) \: d t^2 
+ g^{-1}(z) \: d z^2 
+ e^{A(z) + B(z)} \: d x^2 + e^{A(z) - B(z)} \: d y^2 \right] , 
\label{eqn:metric} \\ 
A_a &=& \phi (z) \frac{L^2}{r_+} \: ( d t )_a \equiv  \Phi (z) \: ( d t )_a . 
\label{eqn:GaugeRS} 
\end{eqnarray}
In what follows, we exclusively use the new coordinate system \eqref{eqn:metric}. 
With the new coordinates, the asymptotic behavior \eqref{eqn:gauge_asympt_r} and \eqref{eqn:scalar_asympt_r} is rewritten as
\begin{eqnarray}
\Phi (z) &\rightarrow & \mu_z -\rho_z \, z  \quad \mathrm{as} \quad z \rightarrow 0 , \label{eqn:gauge_asympt_z}
\\
\Psi (z) &\rightarrow & \Psi_{z1}\:z + \Psi_{z2}\:z^2  
\quad \mathrm{as} \quad z \rightarrow 0 , \label{eqn:scalar_asympt_z}
\end{eqnarray}
where $\mu_z \equiv \frac{L^2}{r_+} \mu$, $\rho_z \equiv \frac{L^2}{r_+^2} \rho$, 
$\Psi_{z1} \equiv \frac{\Psi_1}{r_+}$ and $\Psi_{z2} \equiv \frac{\Psi_2}{r_+^2}$. 
Notice that in our coordinate system the horizon and infinity correspond to $z=1$ and $z=0$ respectively. 
Then, the temperature $T = \kappa / 2 \pi$ is calculated as 
\begin{equation}
T = - \frac{r_+}{4 \pi L^2} g'(1) , 
\end{equation}
where a prime denotes the derivative with respect to $z$. 
We also introduce here the temperature normalized by the charge density $\rho$ as 
\begin{equation}
\normrho{T} \equiv \frac{T}{\sqrt{\rho}} = - \frac{g'(1)}{4\pi L\sqrt{-\Phi' (0)}} , 
\label{eqn:mathcalT} 
\end{equation} 
for later convenience.

For the dilaton field $\varphi$, we make the ansatz
\begin{equation}
\varphi = \alpha \:x,
\end{equation}
where $\alpha$ is a dimensionless constant.
This is a simple way to consider spatial anisotropy in holographic models, 
which was originally used in \cite{MateosTranc2011lett, MateosTranc2011}.
Under this assumption, Eq. \eqref{eqn:scalar} is automatically satisfied. 
We note that this holographic model becomes isotropic if $\alpha$ is zero. 

\subsection{Numerical solution}\label{subsec:NS}
We now explain how to solve the field equations. 
In the coordinate system \eqref{eqn:metric}, Eqs. \eqref{eqn:einstein}--\eqref{eqn:complex} are written as follows
\begin{eqnarray}
 A'' + \frac{1}{2} \left( {A'}^2 + {B'}^2 \right) + {\Psi'}^2 
 + \frac{q^2\Phi^2 \Psi^2}{g^2} &=& 0 , 
\label{eqn:DEA} \\
 B'' + \left( \frac{g'}{g} + A' - \frac{2}{z} \right) B' + \frac{\alpha^2 e^{-(A+B)}}{g} 
 &=& 0 , 
\label{eqn:DEB} \\ 
 \left( A' - \frac{2}{z} \right) g' 
 + \left( \frac{{A'}^2}{2} - \frac{{B'}^2}{2} 
 - {\Psi'}^2 - \frac{4 A'}{z} + \frac{6}{z^2} \right) g
 && \notag \\ 
 - \left( \frac{2}{z^2} + \frac{q^2\Phi ^2}{g} \right) \Psi^2 
+ \frac{z^2\Phi '^2}{2 L^2} - \frac{6}{z^2} + \alpha^2 e^{-(A+B)} &=& 0 , 
\label{eqn:DEConstr} \\  
\Phi'' + A' \Phi' - \frac{2 q^2 L^2\Psi^2}{g z^2}\Phi &=& 0 , 
\label{eqn:DEGauge} \\ 
\Psi'' + \left( \frac{g'}{g} + A' - \frac{2}{z} \right) \Psi' 
+ \left( \frac{2}{g z^2} + \frac{q^2\Phi^2}{g^2} \right) \Psi &=& 0 . 
\label{eqn:DEScalar}
\end{eqnarray}
We note that the equations are invariant under the scale transformation
\begin{equation}
\Phi \rightarrow a\Phi , \quad q \rightarrow q/a , \quad L \rightarrow aL ,
\label{eq:scaling01}
\end{equation}
which rescales the metric \eqref{eqn:metric} by $a^2$.
By employing this rescaling, we set as $L = 1$ in what follows, unless we explicitly restore $L$.

As the boundary condition on the horizon, $z = 1$, we have $g(1) = 0$. 
In addition, we set $\Phi(1) = 0$ so that $A_a A^a$ is finite.  
Then, the regularity of Eqs. \eqref{eqn:DEB}, \eqref{eqn:DEConstr} and \eqref{eqn:DEScalar} yields 
\begin{eqnarray}
g'(1) B'(1) + \alpha^2 e^{- A(1) - B(1)} &=& 0 , 
\label{eqn:DEBH} \\ 
\left( A'(1) - 2 \right) g'(1) - 2 \Psi(1)^2 + \frac{\Phi '^2(1)}{2} + \alpha^2 e^{- A(1) - B(1)} - 6 &=& 0 , 
\label{eqn:DEConstrH} \\
g'(1) \Psi'(1) + 2 \Psi(1) &=& 0 . 
\label{eqn:DEScalarH}
\end{eqnarray}
Therefore, free parameters at the horizon are chosen as
\begin{equation}
A(1) ,\quad B(1) ,\quad \Psi (1) ,\quad \Phi '(1) ,\quad g'(1)
\label{eqn:BCHorizon} 
\end{equation} 
and the rest is determined by Eqs. \eqref{eqn:DEBH}--\eqref{eqn:DEScalarH}.  

On the other hand, we assume that all the variables are sufficiently differentiable and 
$\Psi(z)$ vanishes at infinity. Then, Eq. \eqref{eqn:DEConstr} requires 
\begin{equation}
g(0) = 1 , \quad A'(0) = g'(0) . 
\label{eqn:BCGGA} 
\end{equation} 
Since we are concerned with asymptotically AdS solutions, we impose 
$g'(0) = 0$, or equivalently $A'(0) = 0$. 
The asymptotic solutions to the remaining equations are then derived as 
\begin{equation*}
A (z) = A(0) + \mathrm{O}(z^2) , \quad
B(z) = B(0) + \mathrm{O}(z^2) .
\end{equation*}
In order for the metric Eq. \eqref{eqn:metric} to take the manifestly AdS form, we need to require 
\begin{equation} 
A(0) = 0 , \quad B(0) = 0.  
\label{eqn:BCAB} 
\end{equation} 
When we construct the family of solutions with a \emph{fixed} value of $\alpha$, 
it is not allowed to realize the condition \eqref{eqn:BCAB} by rescaling.
It is because the rescaling of $x$ necessarily involves the rescaling of $\alpha$, 
which results in a family of solutions with different values of $\alpha$. 
Hence, we impose the condition \eqref{eqn:BCAB} as the boundary condition at infinity.

For $\Psi$, both of the two terms in \eqref{eqn:scalar_asympt_z} are normalizable \cite{KlebanovWitten} 
and we can impose the boundary condition that either one vanishes. 
In this paper, we focus on the case where $\Psi_1$ vanishes. 
Furthermore, in the numerical calculations below, we consider solutions with the normalized temperature $\normrho{T}$ fixed. 
We then need to impose the boundary condition at infinity on the gauge potential 
$\Phi$ such that Eq. \eqref{eqn:mathcalT} is satisfied for a fixed $\normrho{T}$. 
Thus, the boundary condition at infinity is summarized as 
\begin{equation}
g'(0) = 0 , \quad A(0) = 0 , \quad B(0) = 0 , \quad \Psi' (0) = 0 , \quad 
\Phi'(0) = - \left( \frac{g'(1)}{4\pi \normrho{T}} \right) ^2 .
\label{eqn:BCInfinity1} 
\end{equation}
Therefore, to obtain numerical solutions of Eqs. \eqref{eqn:DEA}--\eqref{eqn:DEScalar} 
with a fixed temperature $\normrho{T}$, 
we search for the five parameters in Eq. \eqref{eqn:BCHorizon} satisfying 
the boundary condition at infinity Eq. \eqref{eqn:BCInfinity1}, in general.

\subsection{Superconducting state}
\label{subsec:c-temp}
In our holographic model, the charged scalar field $\Psi$ plays the role of the order parameter. 
A normal state of the dual field theory corresponds to the only solution to Eqs. \eqref{eqn:DEA}--\eqref{eqn:DEScalar} 
with $\Psi (z) = 0$, which occurs when the temperature $T$ is higher than a certain temperature $T_c$. 
On the other hand, when $T < T_c$, there emerges a solution with $\Psi \neq 0$, and it corresponds to a superconducting state. 
Therefore, $T_c$ is identified with the critical temperature of the dual field theory. 
Here we analyze the properties associated with the superconducting states, which include the critical temperature, 
the condensate of $\Psi$, and the horizon area.

First we determine the critical temperature. 
To do so, we construct background solutions with $\Psi = 0$ by restricting  
to $\Psi(1) = 0$ in Eq. \eqref{eqn:BCHorizon}, which we have confirmed is 
numerically consistent with the perturbed solution studied in \cite{IizukaMaeda2012}. 
We then consider perturbation on the background solutions and analyze whether a hair, i.e., 
non-trivial configuration of $\Psi$, can reside on it. 
As in the isotropic case \cite{HHHletter2008,HHH2008}, we expect that the superconducting phase transition is second-order, 
as actually confirmed below.    
Then, we define the inifinitesimal parameter $\varepsilon$ as 
\begin{equation*}
\varepsilon = \sqrt{1-T/T_c} 
\end{equation*}
and write $\Psi$, near the phase transition point,
\begin{equation} 
\Psi (z) = \varepsilon \delta \Psi(z) + O (\varepsilon ^2) . 
\end{equation} 
Since the backreaction of $\delta \Psi$ onto the metric and the gauge potential is 
second order in $\varepsilon$, we only have to solve 
\begin{equation}
\delta \Psi'' + \left( \frac{g'}{g} + A' - \frac{2}{z} \right) \delta \Psi' + 
\left( \frac{2}{g \, z^2} + \frac{q^2 \Phi^2}{g^2} \right) \delta \Psi = 0, \label{eq:perturb00}
\end{equation} 
on the background solution with $\Psi = 0$, 
in order to determine the critical temperature $T_c$ ($\varepsilon \rightarrow 0$). 
As we saw in Sec. \ref{subsec:NS}, we impose the boundary condition on $\delta \Psi$ at infinity as  
\begin{equation}
\delta \Psi'(0) = 0 . \label{eq:BCInfinity2}
\end{equation}
For a non-trivial solution of $\delta \Psi$, its value at the horizon, $\delta \Psi(1)$, is non-vanishing but
arbitrary, since Eq. \eqref{eq:perturb00} is linear. 
We then regard Eq. \eqref{eq:perturb00}  as an eigenvalue equation 
with the eigenvalue being $q$. We solve it on a \emph{fixed} background solution, 
and identify the critical temperature as the temperature of the background, and $q$  
is \emph{derived} as the lowest eigenvalue. Since the only free parameter in Eq. \eqref{eq:perturb00} is $q$, 
we note that the critical temperature $T_c$ is determined as a function of $q$. 

\begin{figure}[htbp]
\centering \includegraphics{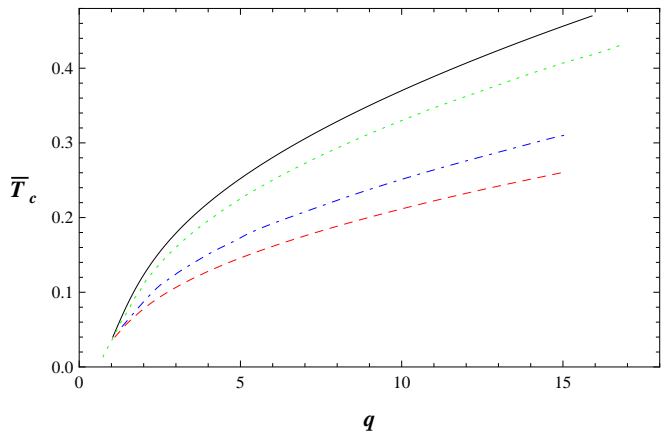}
\caption{
We show $\normrho{T}_c$ as a function of $q$. 
From top to bottom, the various curves correspond to $\alpha = 0$ (black solid line), $1.0$ 
(green dotted line), 
$2.0$ (blue dot-dashed line), $2.5$ (red dashed line).
}
\label{fig:critical_temperature}
\end{figure} 
We show in Fig. \ref{fig:critical_temperature} the critical temperature 
normalized by the charge density, 
$\normrho{T}_c \equiv \left. \normrho{T} \right|_{T = T_c}$, 
for various values of the anisotropic parameter $\alpha$. 
We find that for a fixed $\rho$, $T_c$ monotonically decreases with increasing $\alpha$. 
It has been reported in \cite{MiuraEtAl13} recently that an iron-based superconductor has the same property 
in the sense that anisotropy lowers the critical temperature. 
We note, however, that the anisotropy in \cite{MiuraEtAl13} is between the vertical and parallel directions with respect to the layered superconductor, 
in contrast to our holographic model, where we consider the anisotropy within the layered superconductor (the $x$-$y$ plane).

\begin{figure}[htbp]
\centering \includegraphics{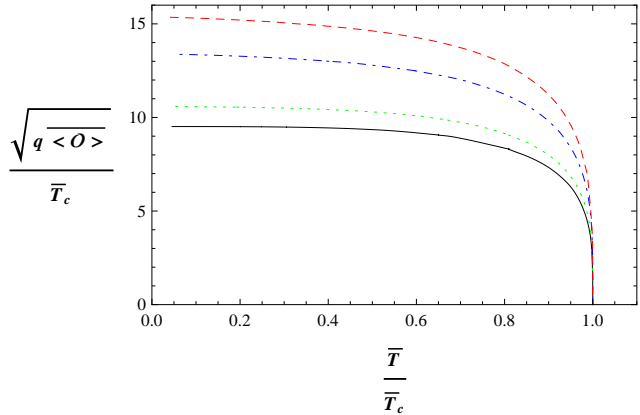}
\caption{
We show the value of the condensate normalized by $\rho / q$, 
$q \normrho{\langle \mathcal{O} \rangle} \equiv q \langle \mathcal{O} \rangle / \rho$, 
further normalized by 
$\normrho{T}_c$, as a function of $\normrho{T} / \normrho{T}_c$ for $q = 3$. 
From bottom to top, the various curves correspond to $\alpha = 0$ (black solid line), $1.0$ (green dotted line), 
$2.0$ (blue dot-dashed line), $2.5$ (red dashed line).
}
\label{fig:condensate}
\end{figure}

Next we investigate the behavior of the condensate of the charged scalar field $\Psi$.
According to the AdS/CFT dictionary, we can read off the value of the condensate $\langle \mathcal{O} \rangle$ as
\begin{equation}
 \langle \mathcal{O} \rangle = \sqrt{2} \Psi_2 , \label{eq:condensate}
\end{equation}
where $\Psi_2$ is defined in Eq. \eqref{eqn:scalar_asympt_r}. 
Thus, now we let $\Psi(1)$ in Eq. \eqref{eqn:BCHorizon} be back to a free parameter, 
and we construct background solutions with $\Psi \neq 0$ by solving Eqs. \eqref{eqn:DEA}--\eqref{eqn:DEScalar}. 
We then determine $\Psi_2$ from their asymptotic forms.
In Fig. \ref{fig:condensate}, we show the value of the condensate as a function of the temperature 
for a variety of the anisotropic parameter $\alpha$. 
Near the phase transition point $T/T_c \approx 1$, the value of the condensate $\langle \mathcal{O} \rangle$ 
is numerically found to behave as
\begin{equation}
 \langle \mathcal{O} \rangle \propto \sqrt{1-T/T_c} ,
\end{equation}
which implies that the phase transition is second-order, as we expected above. 
We see from Fig. \ref{fig:condensate} also that the ratio $\omega_g / T_c$ of the energy gap $\omega_g$ to 
the critical temperature $T_c$ monotonically increases 
with increasing $\alpha$, since the value of the condensate is related to the energy gap of the superconducting state as $\omega_g \simeq 
\sqrt{q \langle \mathcal{O} \rangle}$ \cite{HHHletter2008,HHH2008}.

\begin{figure}[htbp]
\centering \includegraphics{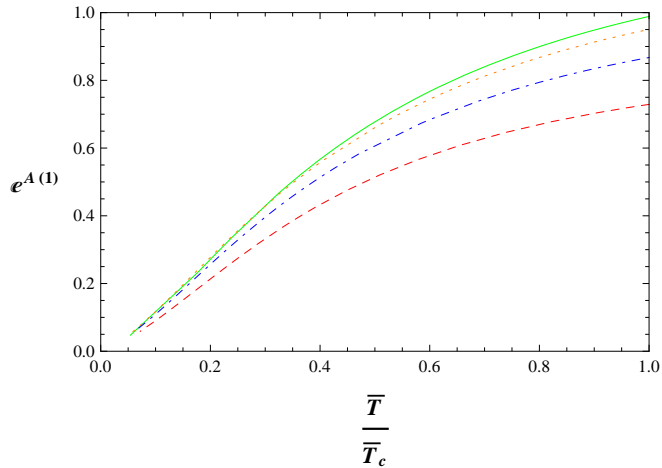}
\caption{
We show the area of the horizon per unit length as a function of $\normrho{T} / \normrho{T}_c$ for $q = 3$. 
From top to bottom, the various curves correspond to $\alpha = 1.0$ (green solid line), $1.5$ (orange dotted line), 
$2.0$ (blue dot-dashed line), $2.5$ (red dashed line).
}
\label{fig:horizon_area}
\end{figure}

It is well-known that the horizon area per unit length of 
the Reissner-Nordstr\"om black brane solution is finite in the extremal limit. 
This feature is not in accord with the Nernst formulation of the third law of thermodynamics and 
hence the solution is not suitable as a gravity dual of holographic superconductor. 
The previous study \cite{IizukaMaeda2012} resolves this problem by considering the normal state in the present holographic model. 
Here we consider whether it remains true also for the superconducting state. 
In our system, the horizon area per unit length is given by 
\begin{equation}
\left. \sqrt{g_{xx}g_{yy}}\right| _{z=1} = e^{A(1)}, 
\end{equation} 
which goes to zero as $T \rightarrow 0$, if the Nernst theorem holds. 
In Fig. \ref{fig:horizon_area}, we show the area in the superconducting state as a function of the temperature. 
We see that the horizon area vanishes in the extremal limit. 
Therefore, our holographic model of superconductor is consistent with the Nernst formulation of the third law of 
thermodynamics also in the superconducting state. 

\section{Conductivity}\label{sec:conductivity}
In this section, we calculate the optical conductivity and investigate the energy gap of the superconducting state. 

\subsection{Perturbation}
In order to calculate the conductivity, we consider the fluctuation of 
the gauge field around the background constructed in Sec. \ref{sec:background}. 
We thus write as
\begin{equation*}
A_a \rightarrow \Phi (z) \: ( dt )_a + \epsilon \left[ \tilde{A}_{x} (t,z) \: ( dx )_a + \tilde{A}_{y} (t,z) \: ( dy )_a \right] ,
\end{equation*}
where $\epsilon$ is the perturbation parameter.
Then, the metric $g_{a b}$ is assumed to be perturbed as
\begin{eqnarray*}
g_{a b} &\rightarrow & \bar{g}_{a b} 
+ 2 \epsilon \frac{L^2}{z^2} \left[ \tilde{g}_{tx} (t,z) \: ( dt dx )_{a b} + \tilde{g}_{ty}(t,z) \: ( dt dy )_{a b} \right] ,
\end{eqnarray*}
where $\bar{g}_{a b}$ is the background metric \eqref{eqn:metric}. 
These perturbation variables are assumed to have the time dependence as 
\begin{eqnarray*}
&&\tilde{A}_{x} (t,z) = e ^{-i \Omega t} A_{x} (z) , \quad
  \tilde{A}_{y} (t,z) = e ^{-i \Omega t} A_{y} (z), \\
&&\tilde{g}_{tx} (t,z) = e ^{-i \Omega t} g_{tx} (z) , \quad
  \tilde{g}_{ty} (t,z) = e ^{-i \Omega t} g_{ty} (z), 
\end{eqnarray*}
where $\Omega$ is a dimensionless constant, with which 
the frequency in the dual field theory is given by 
$\omega = \frac{r_+}{L^2}\Omega$.

If we did not consider perturbation of other degrees of freedom, 
the first-order perturbation of Eq. \eqref{eqn:scalar} would yield
\begin{equation}
\alpha \: \partial _t \tilde{g}_{tx} (t,z) = 0 . \label{eqn:ill-request}
\end{equation} 
This gives $g_{t x} = 0$, and then we obtain $A_x = 0$ from 
Eqs. \eqref{eq:perturb05} below, which implies that the conductivity 
in the $x$-direction is not well-defined. This difficulty is circumvented by adding 
a new degree of freedom of purturbation. Here we assume that the dilaton $\varphi$ is also perturbed as
\begin{equation}
\varphi \rightarrow  \alpha \: x + 
i \, \epsilon \, \Omega \, e ^{-i \Omega t} \, \chi(z). \label{eqn:fluc_dilaton}
\end{equation}

Under the assumptions, the perturbation equations are written as
\begin{eqnarray}
  A_y'' + \left( \frac{g'}{g} + B' \right) A_y' + \left( \frac{\Omega ^2}{g^2} - \frac{2q^2 L^2 \Psi^2}{g z^2} - \frac{z^2\Phi'^2}{gL^2} \right) A_y = 0 ,
 \label{eq:perturb01} \\
  g_{ty}' - \left( A' - B'\right) g_{ty} + \frac{z^2\Phi'}{L^2} A_y = 0 ,
 \label{eq:perturb02} \\
  A_x'' + \left( \frac{g'}{g} - B' \right) A_x' + \left( \frac{\Omega ^2}{g^2} - \frac{2q^2 L^2 \Psi^2}{g z^2} - \frac{z^2\Phi'^2}{gL^2} \right) A_x + 2 \alpha \Phi ' \chi ' = 0 ,
 \label{eq:perturb03} \\
  \chi '' + \left( \frac{g'}{g} + A' -\frac{2}{z} \right) \chi ' + \frac{\Omega ^2}{g^2} \chi - \frac{\alpha e^{-(A+B)}}{g^2} g_{tx} = 0 ,
 \label{eq:perturb04} \\
  g_{tx}' - \left( A' + B'\right) g_{tx} + \frac{z^2\Phi' }{L^2} A_x - 2 \alpha g \chi ' = 0 .
 \label{eq:perturb05}
\end{eqnarray}
Whereas Eq. \eqref{eq:perturb01} is written in a decoupled form, 
we could not decouple Eqs. \eqref{eq:perturb03}--\eqref{eq:perturb05}.    
In Sec. \ref{subsec:NC}, we will transform these perturbation equations into a more practical form.

\subsection{Numerical calculation}\label{subsec:NC}
We note that the perturbation equations \eqref{eq:perturb01}--\eqref{eq:perturb05} 
are linear differential equations having a singular point at the horizon $z=1$ and hence they are numerically unstable. 
Here, we transform Eqs. \eqref{eq:perturb01}--\eqref{eq:perturb05} into a numerically stable form 
and explain how to solve the equations.

We first introduce the new variables as
\begin{equation*}
\hat{A}_y (z) \equiv g(z) \: A_y'(z) , \quad \hat{A}_x(z) \equiv g(z) \: A_x'(z) ,\quad \hat{\chi}(z) \equiv g(z) \: \chi '(z) , 
\end{equation*} 
such that primed and hatted variables possess the same characteristic exponent $\lambda$ for their series expansions
at the horizon. We thus write as 
\begin{align*}
A_y (z) &= (1-z)^{\lambda} \: a_y (z) , & \hat{A}_{y} (z) &= (1-z)^{\lambda} \: \hat{a}_{y} (z), & g_{ty} (z) = (1-z)^{\lambda} \: \zeta _{ty} (z), \\
A_x (z) &= (1-z)^{\lambda} \: a_x (z) , & \hat{A}_{x} (z) &= (1-z)^{\lambda} \: \hat{a}_{x} (z), & g_{tx} (z) = (1-z)^{\lambda} \: \zeta _{tx} (z), \\
\chi (z) &= (1-z)^{\lambda} \: \eta (z) , & \hat{\chi} (z) &= (1-z)^{\lambda} \: \hat{\eta } (z), &
\end{align*} 

In terms of these new variables, we can rewrite the perturbation equations as
\begin{eqnarray}
 \hat{a}_y' - \left( \frac{\lambda}{1-z} - B'\right) \hat{a}_y + \left( \frac{\Omega^2}{g} - \frac{z^2\Phi'^2}{L^2} - \frac{2q^2L^2\Psi^2}{z^2} \right) a_y = 0 ,
 \label{eq:perturb06} \\
 a_y' - \frac{\lambda a_y}{1-z} -\frac{\hat{a}_y}{g} = 0 ,
 \label{eq:perturb07} \\
 \zeta_{ty}' - \left( A'  -B' + \frac{\lambda}{1-z} \right) \zeta_{ty} + \frac{z^2\Phi' a_y}{L^2} = 0 ,
 \label{eq:perturb08} \\
 \hat{a}_x' - \left( \frac{\lambda}{1-z} + B'\right) \hat{a}_x + \left( \frac{\Omega^2}{g} - \frac{z^2\Phi'^2}{L^2} - \frac{2q^2L^2\Psi^2}{z^2} \right) a_x +2\alpha \Phi' \hat{\eta} = 0 ,
 \label{eq:perturb09} \\
 a_x' - \frac{\lambda a_x}{1-z} -\frac{\hat{a}_x}{g} = 0 ,
 \label{eq:perturb10} \\
 \hat{\eta}' - \left( \frac{\lambda}{1-z} -\frac{2}{z} -A' \right) \hat{\eta} + \frac{\Omega^2}{g} \eta - \frac{\alpha e^{-(A+B)}\zeta_{tx}}{g} = 0 ,
 \label{eq:perturb11} \\
 \eta ' - \frac{\lambda \eta}{1-z} -\frac{\hat{\eta}}{g} = 0 ,
 \label{eq:perturb12} \\
 \zeta_{tx}' - \left( A' + B' + \frac{\lambda}{1-z} \right) \zeta_{tx} + \frac{z^2\Phi' a_x}{L^2} -2 \alpha g \hat{\eta} = 0 , 
 \label{eq:perturb13}
\end{eqnarray}
which are written, near the horizon, in the form of the eigenvalue equations as  
\begin{eqnarray}
\begin{bmatrix}
0          & -\Omega ^2 / g'(1) & 0 \\ 
1 / g'(1)  & 0                  & 0 \\
0          & 0                  & 0
\end{bmatrix}
\begin{bmatrix}
\hat{a}_{y} \\ a_{y} \\ \zeta_{ty}
\end{bmatrix}
= \lambda 
\begin{bmatrix}
\hat{a}_{y} \\ a_{y} \\ \zeta_{ty}
\end{bmatrix}
, \label{eq:matrix00}
\\
\begin{bmatrix}
0          & -\Omega ^2 / g'(1)   & 0           & 0                     & 0\\
1 / g'(1)  & 0                    & 0           & 0                     & 0 \\
0          & 0                    & 0           & -\Omega^2 / g'(1)     & \beta / g'(1) \\
0          & 0                    & 1 / g'(1)   & 0                     & 0 \\
0          & 0                    & 0           & 0                     & 0
\end{bmatrix}
\begin{bmatrix}
\hat{a}_x \\ a_x \\ \hat{\eta} \\ \eta \\ \zeta_{tx}
\end{bmatrix}
= \lambda
\begin{bmatrix}
\hat{a}_x \\ a_x \\ \hat{\eta} \\ \eta \\ \zeta_{tx}
\end{bmatrix}
 , \label{eq:matrix01}
\end{eqnarray}
where $\beta \equiv \alpha e^{-A(1)-B(1)}$. 
Here we impose the ingoing boundary condition at the horizon, 
which corresponds to the retarded Green's function in the dual field theory. 
Then the eigenvectors of Eq. \eqref{eq:matrix00} which satisfy this boundary condition are derived as
\begin{eqnarray}
\begin{bmatrix}
0\\0\\1
\end{bmatrix}
, \ \
\begin{bmatrix}
i \Omega \\ 1 \\0
\end{bmatrix}
, \label{eq:eigenvec-y}
\end{eqnarray}
with the corresponding eigenvalues being given by 
$\lambda = 0$ and $\lambda = i\Omega/g'(1)$, respectively. 
We numerically solve Eqs.\eqref{eq:perturb06}--\eqref{eq:perturb08}, by substituting each of these eigenvalues $\lambda$ and taking the corresponding 
eigenvectors in Eq. \eqref{eq:eigenvec-y} as the initial condtion.  
We thus obtain in this way two linearly independent solutions.
Similarly, the eigenvectors of Eq. \eqref{eq:matrix01} satisfying the ingoing boundary condition are found to be
\begin{eqnarray}
\begin{bmatrix}
0\\0\\0\\1\\ \Omega ^2 \beta^{-1}
\end{bmatrix}
, \ \
\begin{bmatrix}
0\\0\\ i \Omega \\ 1 \\ 0
\end{bmatrix}
, \ \
\begin{bmatrix}
i \Omega \\ 1 \\ 0\\0\\0
\end{bmatrix}
, \label{eq:eigenvec-x}
\end{eqnarray} 
where the eigenvalue corresponding to the first eigenvector is 
given by $\lambda = 0$, and the eigenvalues for the other two are found to be 
degenerate and given by $\lambda = i\Omega/g'(1)$. 
Again, we substitute each eigenvalue into Eqs. \eqref{eq:perturb09}--\eqref{eq:perturb13} and 
impose the initial condition as required by the corresponding eigenvectors
in Eq. \eqref{eq:eigenvec-x}, to obtain three linearly independent solutions.

The asymptotic behavior at infinity of the perturbation variables are derived from 
Eqs. \eqref{eq:perturb01}--\eqref{eq:perturb05} as 
\begin{equation} 
g_{t i} = g_{t i}^{(0)} + g_{t i}^{(2)} \: z^2 +\cdots , \quad 
A_i = A_i^{(0)} + A_i^{(1)} \: z +\cdots , \quad 
\chi = \chi^{(0)} + \chi^{(3)} \: z^3 +\cdots , 
\label{eqn:AsyFallOffCond} 
\end{equation} 
where $i$ denotes $x$ or $y$, and $g_{ti}^{(0)}$, $g_{ti}^{(2)}$, $A_i^{(0)}$, $A_i^{(1)}$, $\chi^{(0)}$ and $\chi^{(3)}$ are constants. 
In the AdS/CFT correspondence, a slower fall-off corresponds to a source of the dual field theory. 
In particular, $A_i^{(1)}$ corresponds to the expectation value of the current $\langle J_i \rangle$ of 
the dual field theory for the external electric field $E_i = i\Omega A_i^{(0)}$,  
and then the optical conductivity in the $i$-direction is computed 
\cite{HHHletter2008,HHH2008} as
\begin{equation*}
\sigma_i(\omega) = \frac{\langle J_i \rangle}{E_i} = \frac{A_i^{(1)}}{i \Omega A_i^{(0)}}.
\end{equation*}
On the other hand,
$g_{ti}^{(0)}$ represents a thermal gradient which induces an energy flow in the $i$-direction. 
Here we do not consider such sources except for the gauge potential, and require
\begin{equation}
g_{ty}^{(0)} = g_{tx}^{(0)} = \chi^{(0)} = 0, \label{eq:perturb_condition}
\end{equation}
as the boundary condition at infinity.
The linearly independent solutions constructed as above, i.e., those solutions whose 
initial conditions are given by the eigenvectors in Eq. \eqref{eq:eigenvec-y} and 
Eq. \eqref{eq:eigenvec-x}, respectively, may not satisfy the boundary condition 
Eq. \eqref{eq:perturb_condition}. However, Eqs. \eqref{eq:perturb06}--\eqref{eq:perturb08} 
and Eqs. \eqref{eq:perturb09}--\eqref{eq:perturb13} are linear, and hence we 
superimpose these linearly independent solutions, so that 
Eq. \eqref{eq:perturb_condition} is satisfied.

\subsection{Results} \label{subsec:gap}
Now we show  
the numerical results of the optical conductivity $\sigma_i(\omega)$, 
and analyze the energy gap of our holographic superconductor based on the behavior 
of the optical conductivity. 

\begin{figure}[htbp]
\centering \includegraphics{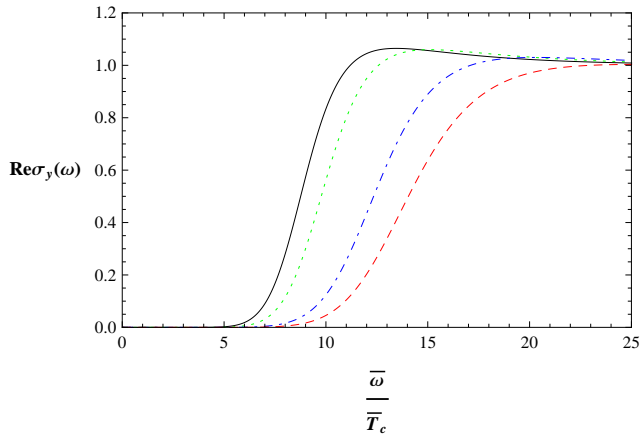}
\caption{
We show $\mathrm{Re} \sigma_y$ as a function of the frequency normalized by 
$\rho$, $\normrho{\omega} \equiv \omega / \sqrt{\rho}$, 
further normalized by $\normrho{T}_c$, for $q = 3$. 
From left to right, the various curves correspond to $\alpha = 0$ (black solid line), $1.0$ (green dotted line), 
$2.0$ (blue dot-dashed line), $2.5$ (red dashed line).
This figure shows that the energy gap increases with increasing $\alpha$. 
}
\label{fig:y-gap1}
\end{figure}

We first consider the conductivity in the $y$-direction. 
We show in Fig. \ref{fig:y-gap1} the real part of conductivity in the $y$-direction. 
Each curve in Fig. \ref{fig:y-gap1} is the limit curve, to which   
the conductivity curves converge as the temperature is lowered. 
We see from Fig. \ref{fig:y-gap1} that the conductivity in the $y$-direction exhibits 
behavior similar to the isotropic case \cite{HHHletter2008,HHH2008}. 
In particular, we find that $\mathrm{Re} \sigma_y(\omega)$ drops to zero 
as the frequency $\omega$ decreases, and hence that the energy gap is well-defined 
in this case. 

\begin{figure}[htbp]
\centering \includegraphics{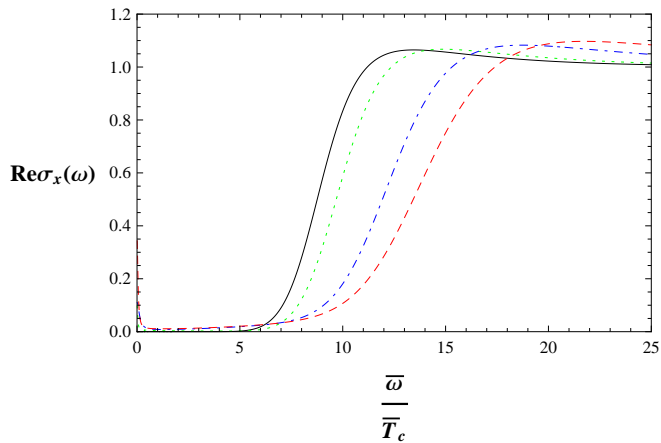}
\caption{
We show $\mathrm{Re} \sigma_x$ as a function of $\normrho{\omega}$ normalized by $\normrho{T}_c$ for $q = 3$.  
From left to right, the various curves correspond to $\alpha = 0$ (black solid line), $1.0$ (green dotted line), 
$2.0$ (blue dot-dashed line), $2.5$ (red dashed line).
}
\label{fig:x-gap}
\end{figure}

In contrast, the conductivity in the $x$-direction shows rather different behavior even for the sufficiently low temperature, 
as shown in Fig. \ref{fig:x-gap}.  
We particularly find that $\mathrm{Re} \sigma_x(\omega)$ remains non-vanishing 
even in the low frequency region. This feature of the conductivity is very similar to 
the psuedogap observed in the p-wave superconductor \cite{GubserPufu2008, RobertsHartnoll2008, HutasoitSiopsisTherrien2014}. 
Closer observation of Fig. \ref{fig:x-gap} reveals also 
that the conductivity within the psuedogap becomes large for the increasing values of the anisotropy parameter $\alpha$.
Since the psuedogap appears in the conductivity in the $x$-direction, 
we estimate the energy gap of our holographic superconductor based only on the conductivity in the $y$-direction. 
For simplicity, in this paper, we shall define the energy gap $\omega_g$ as 
the frequency at the inflection point of the conductivity curve. 
As in the isotropic case \cite{HHH2008}, $\mathrm{Re} \sigma_y$ will get steeper for the larger value of $q$, and then the energy gap 
will not be so sensitive to $q$. Within this accuracy, we find that the energy gap $\omega_g$ is approximately estimated as   
\begin{equation}
\omega_g \simeq (8.1 + 2.0 \alpha) \: T_c . \label{eqn:gap}
\end{equation}
As in the isotropic case \cite{HHH2008}, $\omega_g / T_c$ is found to be large enough 
compared to the one predicted by BCS theory, and hence this feature remains true also in the anisotropic case. 

\begin{figure}[htbp]
\centering \includegraphics{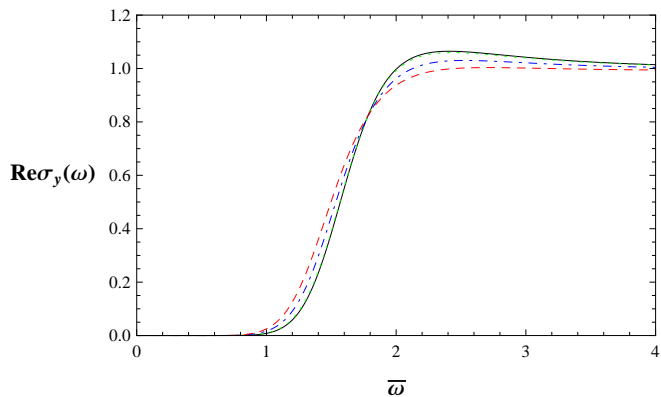}
\caption{
We show $\mathrm{Re} \sigma_y$ as a function of $\normrho{\omega}$ for $q = 3$. 
From left to right, the various curves correspond to $\alpha = 0$ (black solid line), $1.0$ (green dotted line), 
$2.0$ (blue dot-dashed line), $2.5$ (red dashed line). 
}
\label{fig:y-gap2}
\end{figure}

As we saw in Sec. \ref{subsec:c-temp}, the critical temperature $T_c$ depends on the anisotropic parameter $\alpha$. 
It may be helpful to take into account this $\alpha$ dependence of $T_c$ for 
understanding of the total dependence of $\omega_g$ on $\alpha$. 
As we see from Fig. \ref{fig:y-gap2}, the curves of $\mathrm{Re} \sigma_y$ are almost 
degenerate for various values of $\alpha$. Thus, we find that the energy gap itself is insensitive to the anisotropy parameter $\alpha$.

\section{Discussion}\label{sec:discussion}

In this paper, we considered the anisotropic holographic superconductors 
in the Einstein-Maxwell-dilaton theory. 
We first constructed numerically the bulk background black brane solutions corresponding 
not only to the normal state, but also to the superconducting state. 
We then calculated the critical temperature $T_c$, and found that it decreases as the anisotropy becomes large. 
It is very interesting to note that a real-world iron-based superconductor with the same property 
has been reported recently \cite{MiuraEtAl13}. 

We also computed the optical conductivities in both of the $x$- and $y$-directions. 
We estimated the energy gap $\omega_g$ from the conductivity in the $y$-direction and found that $\omega_g / T_c$
increases as the anisotropy becomes large, and hence that $\omega_g$ is larger than $8 T_c$. 
As it is much larger than the BCS prediction $3.54T_c$, the strong coupling effect for the holographic superconductor 
model is not broken by the anisotropy, but rather enhanced.

On the other hand, the conductivity in the $x$-direction never drops to zero even 
at the sufficiently low temperature. This indicates that a pseudogap appears much like in 
the p-wave superconductors \cite{GubserPufu2008, RobertsHartnoll2008, HutasoitSiopsisTherrien2014}, 
and we found that the magnitude of the pseudogap seems to increase 
as the anisotropy becomes large. 
It will be interesting if a peudogap appears generally in an anisotropic model of holographic superconductors, without depending on details of holographic models.

\acknowledgments 
We are grateful to T. Okamura for helpful discussion. We also thank M. Miura for 
enlightening us on the experiment \cite{MiuraEtAl13}.
The work of K.~M. was supported in part by JSPS KAKENHI Grant Number 23740200.

\end{document}